\documentclass[3p,times]{elsarticle}

\usepackage{ecrc}

\volume{00}

\firstpage{1}

\journalname{Physics Procedia}

\runauth{W. Kob {\it et al.}}

\jid{phpro}

\jnltitlelogo{Physics Procedia}

\usepackage{amssymb}

\usepackage[figuresright]{rotating}

\begin{document}

\begin{frontmatter}



\dochead{}

\title{Spatial correlations in glass-forming
liquids across the mode-coupling crossover}


\author{Walter Kob}
\address{Laboratoire Charles Coulomb,
UMR CNRS 5221, Universit{\'e} Montpellier 2,
34095 Montpellier, France}

\author{S\'andalo Rold\'an-Vargas}
\address{Laboratoire Charles Coulomb,
UMR CNRS 5221, Universit{\'e} Montpellier 2,
34095 Montpellier, France}
\address{Departamento de F\'{i}sica
Aplicada, Grupo de F\'{i}sica de Fluidos y Biocoloides, Universidad de
Granada, 18071 Granada, Spain}

\author{Ludovic Berthier}
\address{Laboratoire Charles Coulomb,
UMR CNRS 5221, Universit{\'e} Montpellier 2,
34095 Montpellier, France}

\begin{abstract}
We discuss a novel approach, the point-to-set correlation functions,
that allows to determine relevant static and dynamic length scales in
glass-forming liquids.  We find that static length scales increase
monotonically when the temperature is lowered, whereas the measured
dynamic length scale shows a maximum around the critical temperature of
mode-coupling theory. We show that a similar non-monotonicity is found in
the temperature evolution of certain finite size effects in the relaxation
dynamics. These two independent sets of results demonstrate the existence
of a change in the transport mechanism when the glass-former is cooled
from moderately to deeply supercooled states across the mode-coupling
crossover and clarify  the status of the theoretical calculations done
at the mean field level.
\end{abstract}


\end{frontmatter}



\section{Introduction}

{Within} the last 15 years intensive investigations have shown that
glass-forming systems have a heterogeneous relaxation dynamics, i.e. that
at any given time there are regions in space in which the dynamics is
{significantly} faster than the average dynamics and other regions in
which it is slower~\cite{kob_97,ediger_00,berthier_04}. Although at the
beginning these dynamical heterogeneities where considered to be just an
oddity of glass-formers, it has subsequently been understood that they
are likely to hold an important key to our understanding for the slowing
down of the relaxation dynamics in such systems.  This insight has led
to a very strong research activity in which the properties of dynamic
heterogeneities have been investigated in great detail~\cite{book}.  These
studies have given evidence that the number of particles participating
in the collective motion associated with dynamic heterogeneity increases
with decreasing temperature from $O(1)$ to $O(10^3)$ if $T$ is lowered
from normal liquid temperatures down to $T_g$~\cite{berthier_05,dalle_07}.
As a consequence these domains can be characterized by a {\it dynamical}
length scale that grows if $T$ is lowered~\cite{book,binder_kob_11}.

This growth is in qualitative agreement with mean field theoretical
calculations~\cite{BBepl,silvio07}, which show that there is indeed a
dynamical length scale that is predicted to diverge at $T_c$, the critical
temperature of the mode-coupling theory~\cite{gotze_08}, according to
$\xi^{\rm dyn} \propto (T-T_c)^{-0.25}$~\cite{silvio07,imct}.  However,
this prediction seems to be at odds with the results from experiments
that do not indicate any divergence of the dynamical length scale
around $T_c$~\cite{berthier_05,dalle_07}.  This discrepancy might
be due to the fact that in experiments the dynamical length scale is
determined only in a rather indirect manner, and therefore it is not
evident that in real systems the dynamical length scale does indeed not
show any particular signature around $T_c$. Another possibility could
be that this prediction is not relevant outside the realm of mean field
approximations and that mode-coupling calculations are of little use to
interpret dynamic heterogeneities.

\begin{figure}
\includegraphics[width=0.4\textwidth, bb=-40mm 0mm 70mm 130mm]{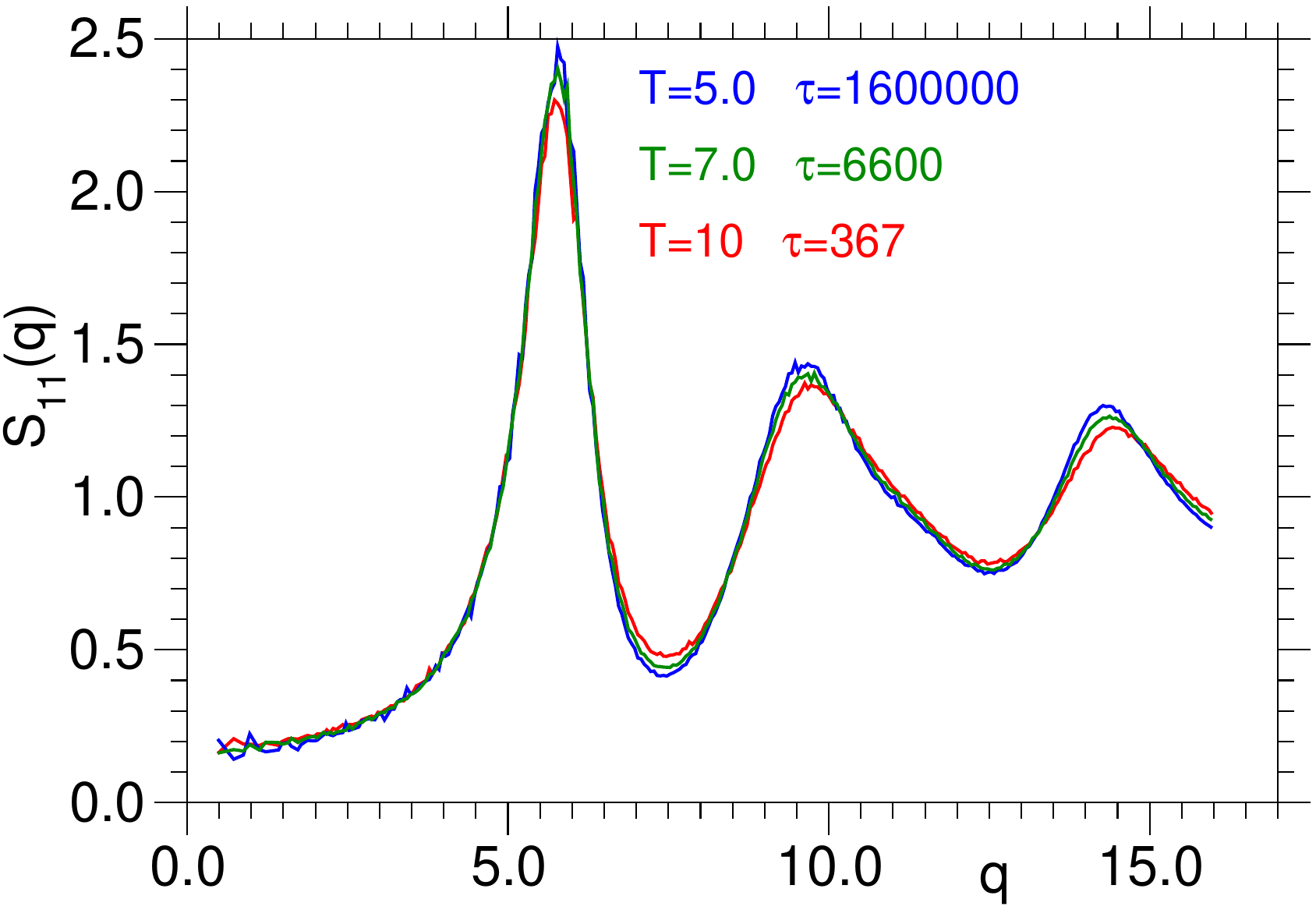}
\caption{Partial structure factor as a function of wave-vector $q$
for the binary system studied in this work. The three curves correspond to the
temperatures given in the legend. Also given are the $\alpha-$relaxation
times $\tau$ of the system at the corresponding temperatures.}
\label{fig1}
\end{figure}

Another important question regarding the dynamical heterogeneities
concerns the microscopic mechanism which is responsible for their
existence: Are they just thermal fluctuations of the particle density
or of the velocities, or {do they instead reflect the emergence of}
some `special' amorphous structures that become more extended when
temperature is decreased? Experiments and simulations indicate that
two-point correlation functions, such as the radial distribution function
$g(r)$ or the static structure factor $S(q)$, do not show any remarkable
temperature dependence, in contrast to what is found in the vicinity of
second order phase transitions. As an example we show in Fig.~\ref{fig1}
the partial structure factor of a glass-forming system (described below)
at three different temperatures, demonstrating that $S(q)$ only shows a
mild temperature dependence.  This is in contrast with the $T-$dependence
of the relaxation times $\tau$ (which can, e.g., be determined from the
decay time of the intermediate scattering function), that change in the
same $T-$range by a factor of about $10^4$.

Since two-point correlation functions do not show a significant
$T-$dependence one is lead to probe higher order correlation
functions. Such a line of research is also motivated by the fact that
the so-called random first order transition (RFOT) theory of the glass
transition predicts that glass-formers do show at low temperatures a
{\it static} order~\cite{kirpatrick_89,xia_00}. Within this theory, this
order becomes relevant around the mode-coupling temperature $T_c$ and,
within mean field, is found to diverge at the ideal glass transition,
also called Kauzmann temperature $T_K < T_c$.

Although computer simulations have given evidence that
certain structural motifs do indeed become more frequent at
low temperatures~\cite{coslovich_07,tanaka}, it is at present
not clear at all whether these locally favored structures are
only relevant in the moderately supercooled regime or also
at temperatures close to the glass transition~\cite{gilles}.
As a consequence there have recently been significant efforts to
develop methods to define and quantify such increasing static length
scales~\cite{bb_04,jack_05,cavagna_07,biroli_08,mosayebi_10,kob_12,berthier_12,jack_12,charbonneau_12}.
These studies are rendered difficult by the fact that the length scales
are expected to become resonably large only at temperatures around $T_c$,
a temperature at which the dynamics of the glass-former is usually very
sluggish. Hence current computer simulations have a hard time to probe
the {\it equilibrium} dynamics of the liquid in this temperature range.
Nevertheless, these efforts are highly justified since one needs to
clarify how these two strong theoretical predictions, that the dynamic
length scales diverges at $T_c$ whereas the static scale diverges only
at $T_K< T_c$, compare with measurements in real systems. We want to
determine if and how these two seemingly contradictory results survive
in real glass-formers.  In the following we will therefore present the
results of large scale computer simulations in which we have addressed
the question on the presence of dynamical and static length scales,
how they depend on temperature, and how they are related to each other.

\section{Model and Details of the Simulations}

The model we are using is a 50:50 binary mixture of harmonic spheres
with different sizes~\cite{berthier_09}. All particles have the same
mass $m$ and particles $i$ and $j$ that are a distance $r_{ij}$ apart
have an interaction given by

\begin{equation}
V(r_{ij}) = \frac{\epsilon}{2} (1-(r_{ij}/\sigma_{ij})^2) \quad .
\label{eq1}
\end{equation}

\noindent
Here $\sigma_{11}=1.0$, $\sigma_{12}=1.2$ and $\sigma_{22}=1.4$. In
the following we will use $\sigma_{11}$ as the unit of distance,
$\sqrt{m\sigma_{11}/\epsilon}$ as the unit of time and $10^{-4}\epsilon$
as unit of energy, setting the Boltzmann constant $k_B=1.0$. The equations
of motion are integrated using the velocity form of the Verlet algorithm,
using a time step of 0.02. We use two distinct protocols to study
this system.

In a first study~\cite{kob_12}, we have equilibrated a system of 4320
particles using a box of size $L_x=L_y=13.68$ and $L_z=34.2$. After
equilibration, we have permanently frozen the position of all the
particles that had a coordinate $-1.4 \leq z \leq 0$ (using periodic
boundary conditions) and introduced at $z=0$ a hard wall.  Therefore
these particles become a frozen amorphous wall. {It can be shown that
this freezing does {not} influence the static properties of the remaining
fluid particles~\cite{scheidler}}.  We have performed a disorder average
over 10-30 independent configurations of the wall. Due to the presence
of an amorphous wall, {this setup allows us to study how the static
and dynamic properties of the liquid are influenced by the presence
of the frozen particles}. Since the latter occupy positions that are
characteristic of the bulk liquid, this point-to-set approach probes
static and dynamic bulk correlation functions. In the following we will
characterize the static and dynamic properties of the system as a function
of the distance $z$ from the frozen wall.

In a second set of simulations~\cite{berthier_tobe}, we have performed
equilibrium simulations of the bulk system using various system sizes
$N$ from $N=32$ up to $N=16384$ keeping the density constant.  For each
temperature and system size we determine the bulk relaxation time,
$\tau(L,T)$, and study at each temperature the effect of decreasing the
system size on the bulk relaxation.  Our goal is to study the interplay
between a finite system size and the relevant correlation length scales
characterizing the glass-former in the bulk.

\section{Results}

In order to characterize the static and dynamic properties of the
system we introduce a coarse-grained density field in the following
manner~\cite{biroli_08}. We decompose the simulation box into small
non-overlapping boxes (cells) of size $\delta \approx 0.55$ and define a
binary variable $n_i(t)=1$ if at time $t$ cell number $i$ is occupied by
at least one particle, and $n_i(t)=0$ if not. From this coarse grained
density we can now define a time correlation function, also called
`overlap',

\begin{equation}
q_{c}(t,z)=\left[\frac{\sum_{i(z)} \langle n_i(t) n_i(0)
\rangle}{\sum_{i(z)} \langle n_i(0)\rangle}\right]_{\rm wall} \quad ,
\label{eq2}
\end{equation}

\noindent
which gives the probability that if a cell has been occupied at time
$t=0$ it is also occupied at time $t$. Note that in Eq.~(\ref{eq2})
the sum extends over all cells that are at distance $z$ from the wall
and $[ \cdots]_{\rm wall}$ denotes the average over the independent
walls. We note that $q_{c}(t,z)$ is a collective quantity in {the sense}
that we do not specify in its definition the {label} of the particle
that occupies a given cell. It is, however, also possible to define a
`self-overlap', $q_{s}(t,z)$, by requiring that at times $t=0$ and $t$
a given cell is occupied by the {same} particle and below we will discuss
this correlator as well.

\begin{figure}[t]
\includegraphics[width=35mm, bb=-20mm 60mm 100mm 170mm]{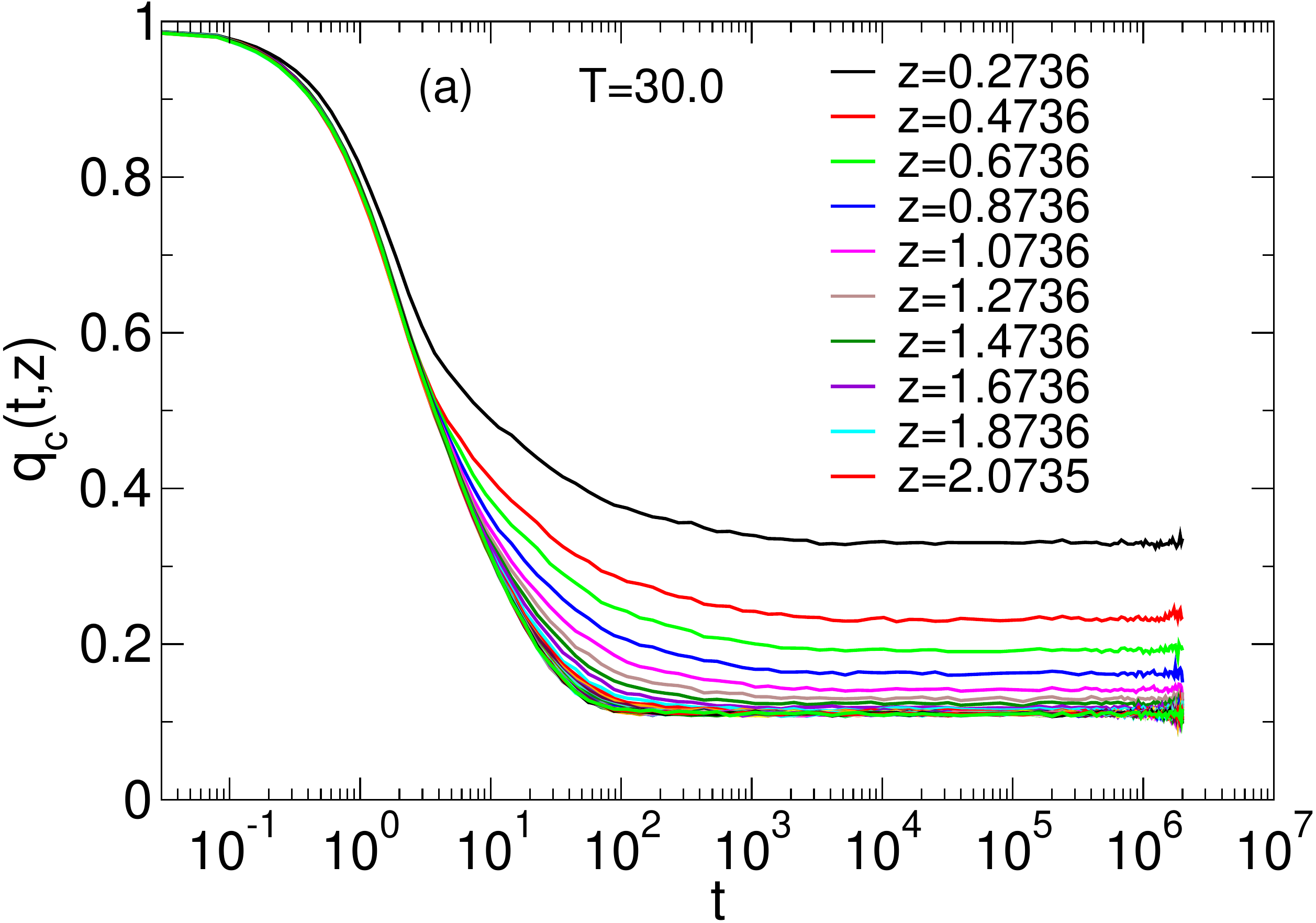}
\\[-36mm]

\hspace*{75mm}
\includegraphics[width=35mm,bb=-20mm 60mm 100mm 170mm]{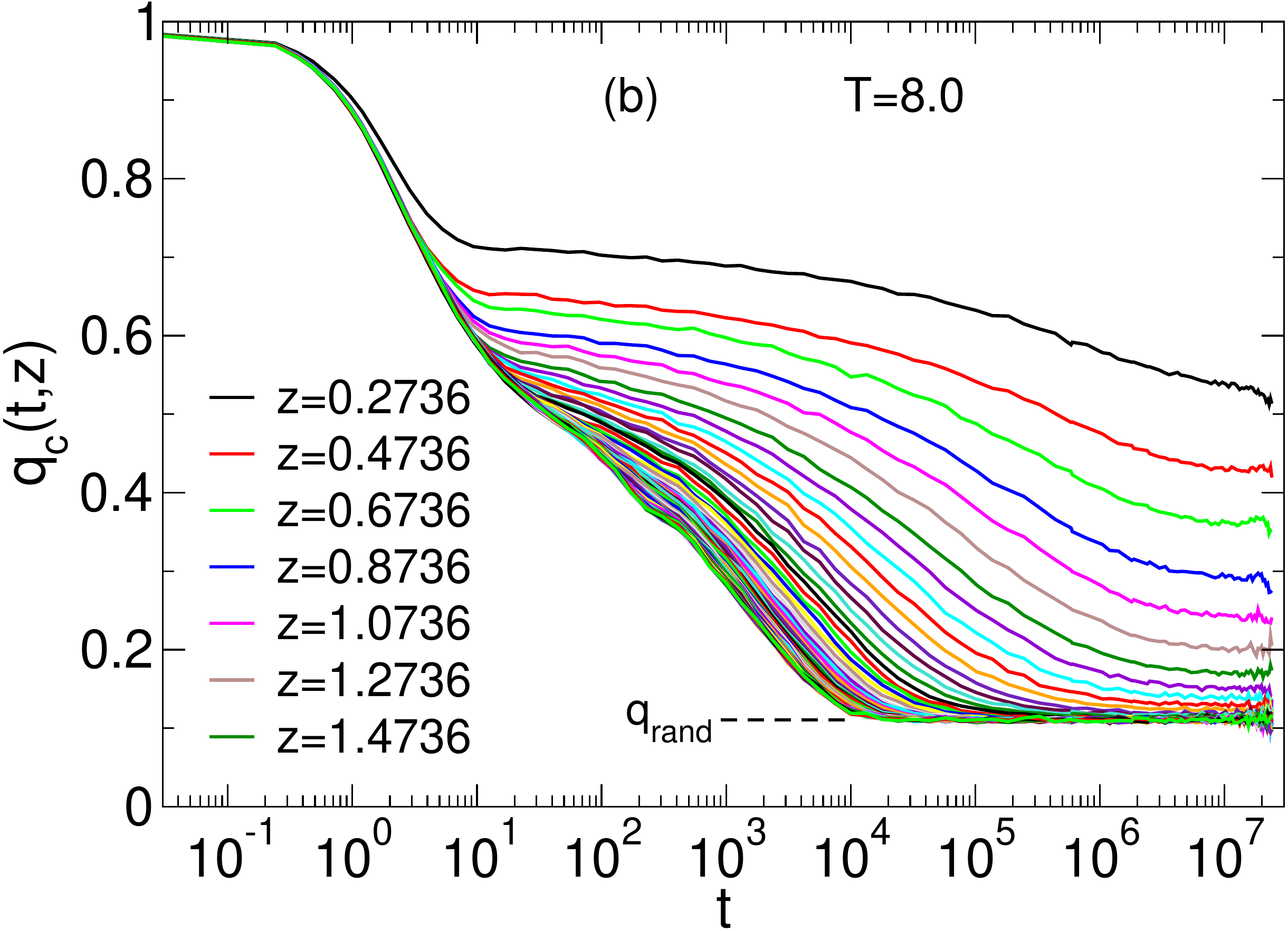}
\\[18mm]

\includegraphics[width=35mm, bb=-20mm 60mm 100mm 170mm]{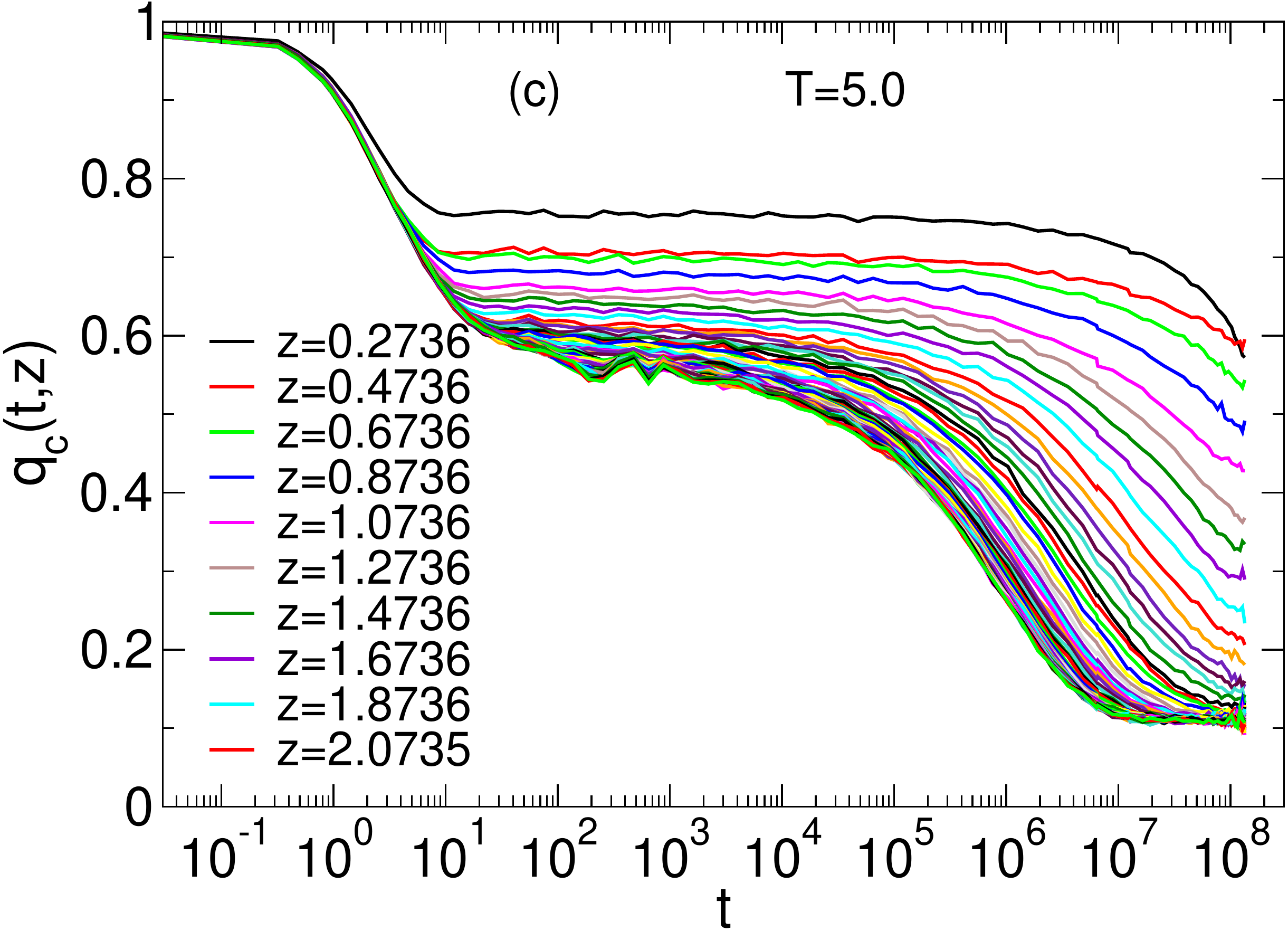}
\\[9mm]

\caption{Time dependence of the correlation function $q_{c}(t,z)$ as defined
in Eq.~(\ref{eq2}) for different values of $z$. The three panels correspond
to the temperatures given in the figures.}
\label{fig2}
\end{figure}

In Fig.~\ref{fig2} we show the time dependence of $q_c(t,z)$ for different
values of $z$. The three panels correspond to different temperatures.
We see that at high temperatures (Fig.~\ref{fig2}a) $q_c(t,z)$ decays
quite quickly to a $z-$dependent constant. This decay is very fast
for intermediate and large values of $z$, whereas it is slower for
small values of $z$. Thus we see that at this temperature the density
correlation function relaxes quickly and becomes independent of $z$
{for large distances} from the frozen wall and that the relaxation
dynamics slows down if the wall is approached, in agreement with previous
findings~\cite{scheidler}.  If $T$ is decreased to a temperature at
which the relaxation dynamics of the bulk system starts to become
somewhat sluggish~\cite{berthier_09} one finds qualitatively the same
results (Fig.~\ref{fig2}b). The main difference is that for this $T$ the
correlation function shows for all values of $z$ at intermediate times
a plateau, a feature that is well known from the relaxation dynamics in
the bulk and that corresponds to the temporary trapping of the particles
inside the cage formed by their neighbors~\cite{binder_kob_11}. From
the figure we recognize that the width of this plateau increases
with decreasing $z$, i.e. the trapping becomes more pronounced if one
approaches the wall. This effect is related to the fact that the particles
constituting the walls do not have any thermal motion and hence do not
provide any fluctuations that can help the particles inside the liquid
to relax. This effect becomes even stronger if temperature is lowered
further, Fig.~\ref{fig2}c. At this temperature, which is somewhat below
the critical temperature of mode-coupling theory~\cite{kob_12}, the
relaxation dynamics close to the wall is so slow that it is not possible
to see the final relaxation within the time window of our simulation. This
does not imply that our simulations are out of equilibrium since, by
construction of the wall, we have started our production runs with an
equilibrated state.  It implies, however, that we cannot accurately
extract the long-time limit of these overlap functions for all values
of $z$.

\begin{figure}[t]
\hspace*{-7mm}
\includegraphics[width=65mm, bb=-40mm -20mm 90mm 120mm]{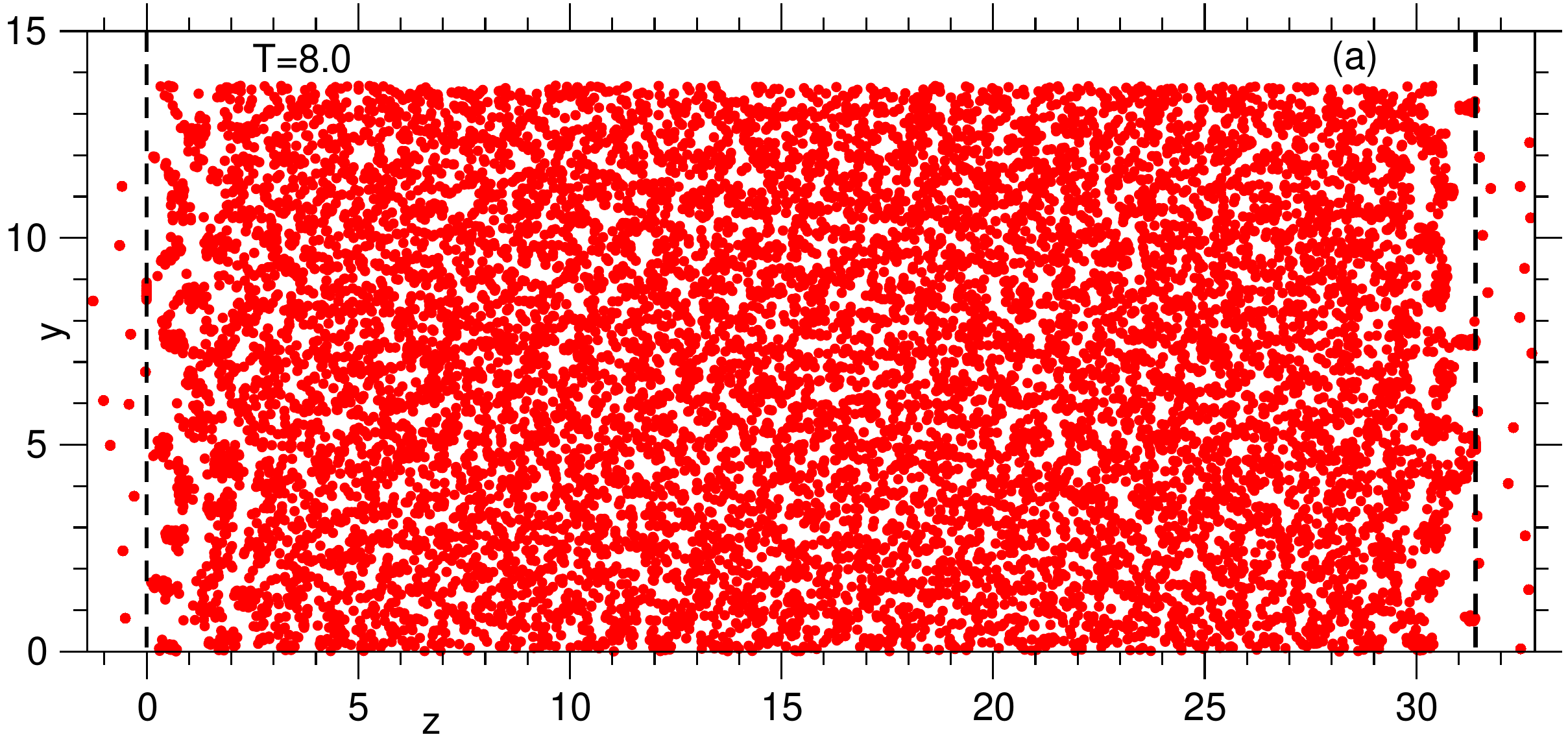}
\\[-22mm]

\hspace*{-7mm}
\includegraphics[width=65mm, bb=-40mm -20mm 90mm 120mm]{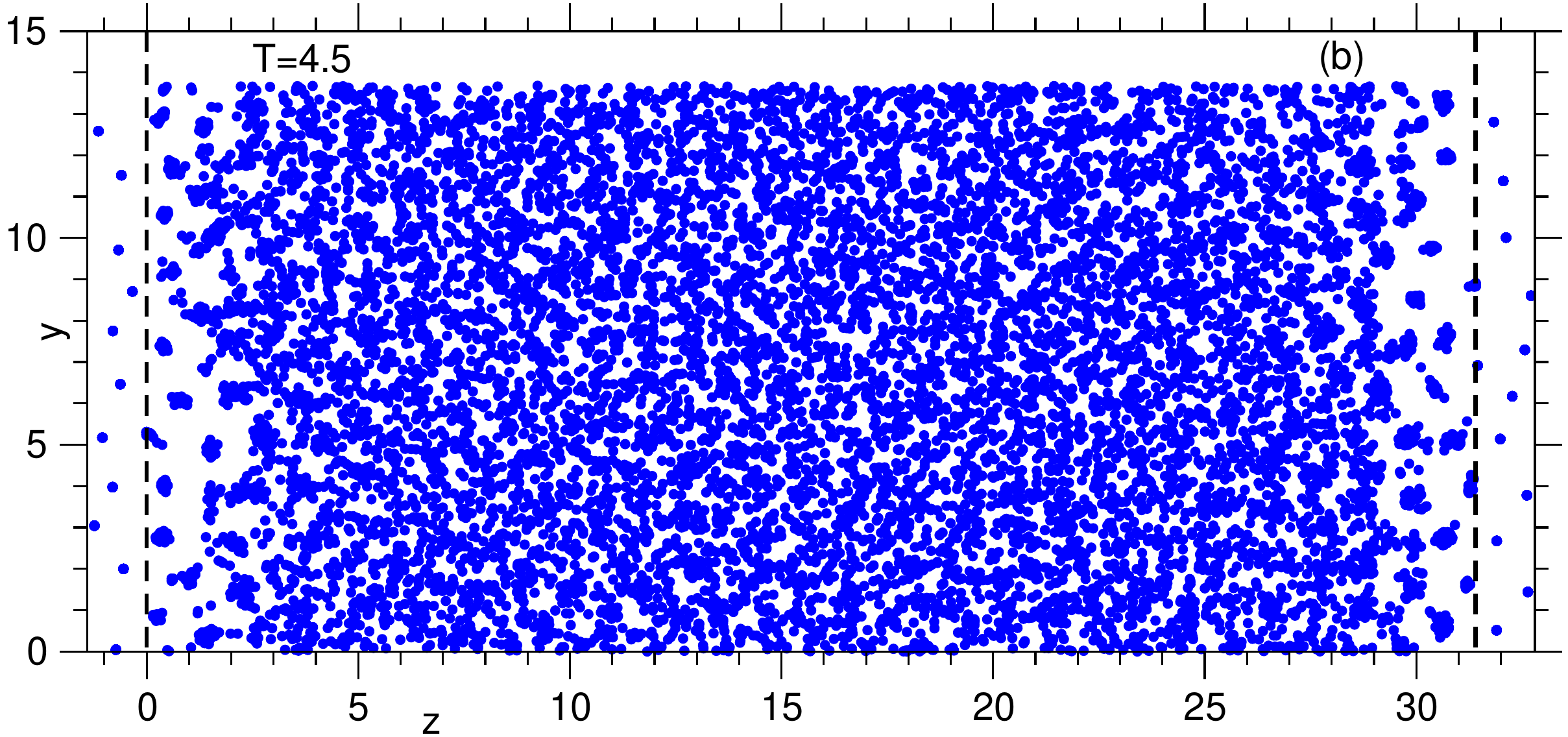}
\vspace*{-12mm}
\caption{Superposition of snapshots of the system at $T = 8.0$ and $T = 4.5$
(panel (a) and (b), respectively). The vertical dashed lines show the
location of the hard walls. Only particles in the slab $0\leq  x \leq
1.0$ are shown. Individual snapshots are separated by about $\tau_{\rm
bulk}(T)$ and the total time is about $100 \tau_{\rm bulk}(T)$.}
\label{fig3}
\end{figure}

From Fig.~\ref{fig2} we see that at intermediate and long times
$q_{c}(t,z)$ can be approximated by the functional form

\begin{equation}
q_c(t,z) = A(z,T) \exp[-(t/\tau_c(z,T))^{\beta(z,T)}]+q_\infty(z,T)+
q_{\rm rand}.
\label{eq3}
\end{equation}

\noindent
Here $q_{\rm rand}$ is the long time value of $q_{c}(t,z)$ for large
$z$, i.e. it is a quantity that can be obtained with high precision
from a simulation of the bulk. We find that $q_{\rm rand} \approx
0.110595$, independently of $T$.  Therefore, the quantity $q_\infty(z,T)$
quantifies the nontrivial influence of the wall on the {local} density
field. Qualitatively this influence can be seen in Fig.~\ref{fig3} where
we show the superposition of many particle configurations obtained during
a run that extended over about 100 times the typical relaxation time
of the bulk system. In order to avoid overcrowding we show only a thin
slice of the system, i.e. the particles that have coordinates $0\leq x
\leq 1.0$. From these figures we see that {far from the wall the system
is fully ergodic, and particles fill out all the available space in a
uniform manner. By contrast, the space close to the wall is occupied in a
very heterogeneous manner, exhibiting locations in which there is a high
probability to find a particle and others in which this probability is
small}. A comparison of the two panels shows that the length scale over
which the wall influences the local structure increases if temperature
is decreased, thus giving rise to an increasing static length scale.

We find that the self-overlap $q_s(z,t)$ shows a qualitatively similar
time dependence as $q_c(z,t)$. The main difference is that at long times
the former always decays to zero since the probability that a given cell
is occupied at very long times by the particle that was present at time
zero becomes negligibly small. Therefore the quantity $q_{\infty}(z,T)$,
now associated to $q_s(z,t)$, is zero for all values of $z$ and $T$.

\begin{figure}[t]
\hspace*{30mm}
\includegraphics[width=90mm]{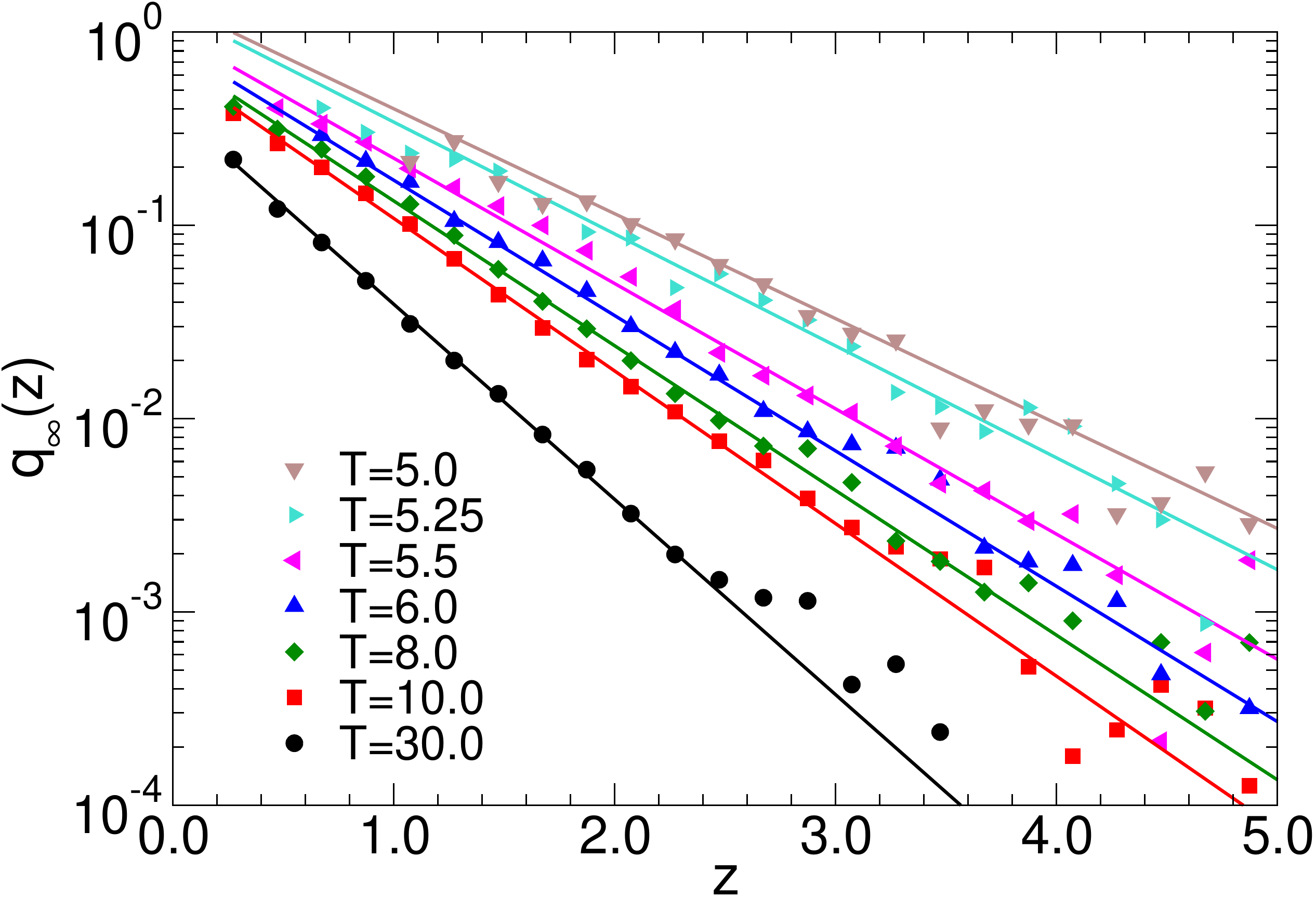}
\caption{$z-$dependence of the long-time overlap for different
temperatures (symbols). The straight lines are fits with an exponential.}
\label{fig4}
\end{figure}

The increasing influence of the wall on the local structure of the
liquid can be quantified by the static overlap $q_\infty(z,T)$ from
Eq.~(\ref{eq3}), which gives the excess probability to find a particle
in a given cell with respect to the bulk probability. This quantity
can be obtained with quite high precision by fitting the final decay
of $q_c(t,z)$ with the functional form given in Eq.~(\ref{eq3}) and in
Ref.~\cite{kob_12} we have shown that this expression does indeed give
a very good description to the data.  The resulting $z-$dependence
of $q_\infty(z,T)$ is shown in Fig.~\ref{fig4}. We see that for
all temperatures investigated $q_\infty(z,T)$ is compatible with an
exponential dependence in $z$ (straight lines):

\begin{equation}
q_{\infty}(T)=B(T) \exp(-z/\xi^{\rm stat}(T)).
\label{eq4}
\end{equation}

\noindent
This functional form allows us thus to define a $T-$dependent
``point-to-set'' length scale $\xi^{\rm stat}$ that grows if the
temperature is decreased. Note that the data points imply that not only
$\xi^{\rm stat}$ depends on $T$ but the prefactor $B$ from Eq.~(\ref{eq4})
does as well since the curves move upwards with decreasing $T$. This
suggests to define a second static length scale $\xi^{\rm stat-int}$
via the integral of $q_{\infty}(T)$ which, with Eq.~(\ref{eq4}), leads
to the estimate

\begin{equation}
\xi^{\rm stat-int} \approx B(T) \cdot \xi^{\rm stat}.
\label{eq4b}
\end{equation}

\begin{figure}[b!]
\includegraphics[scale=0.32,angle=0,clip]{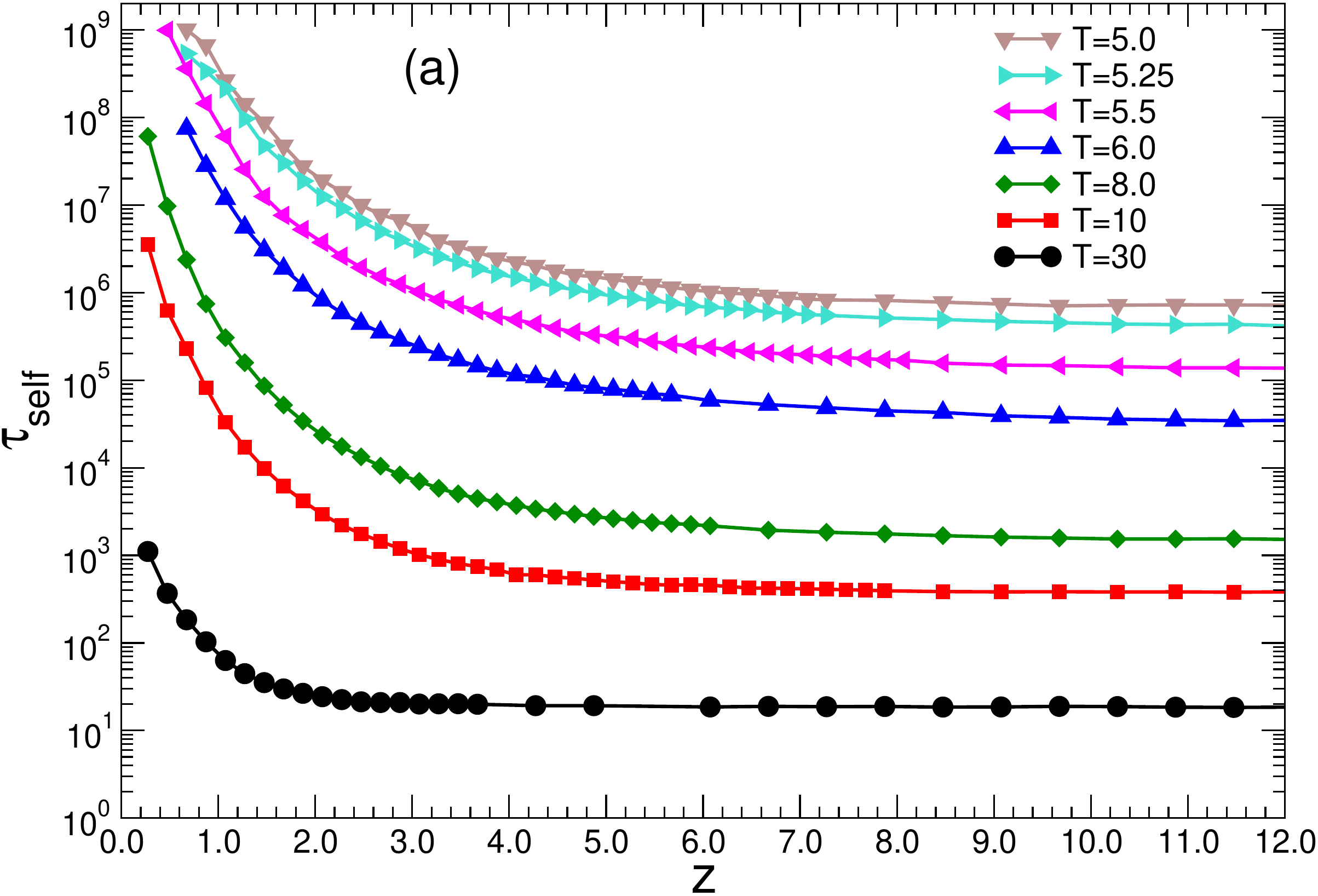}
\includegraphics[scale=0.32,angle=0,clip]{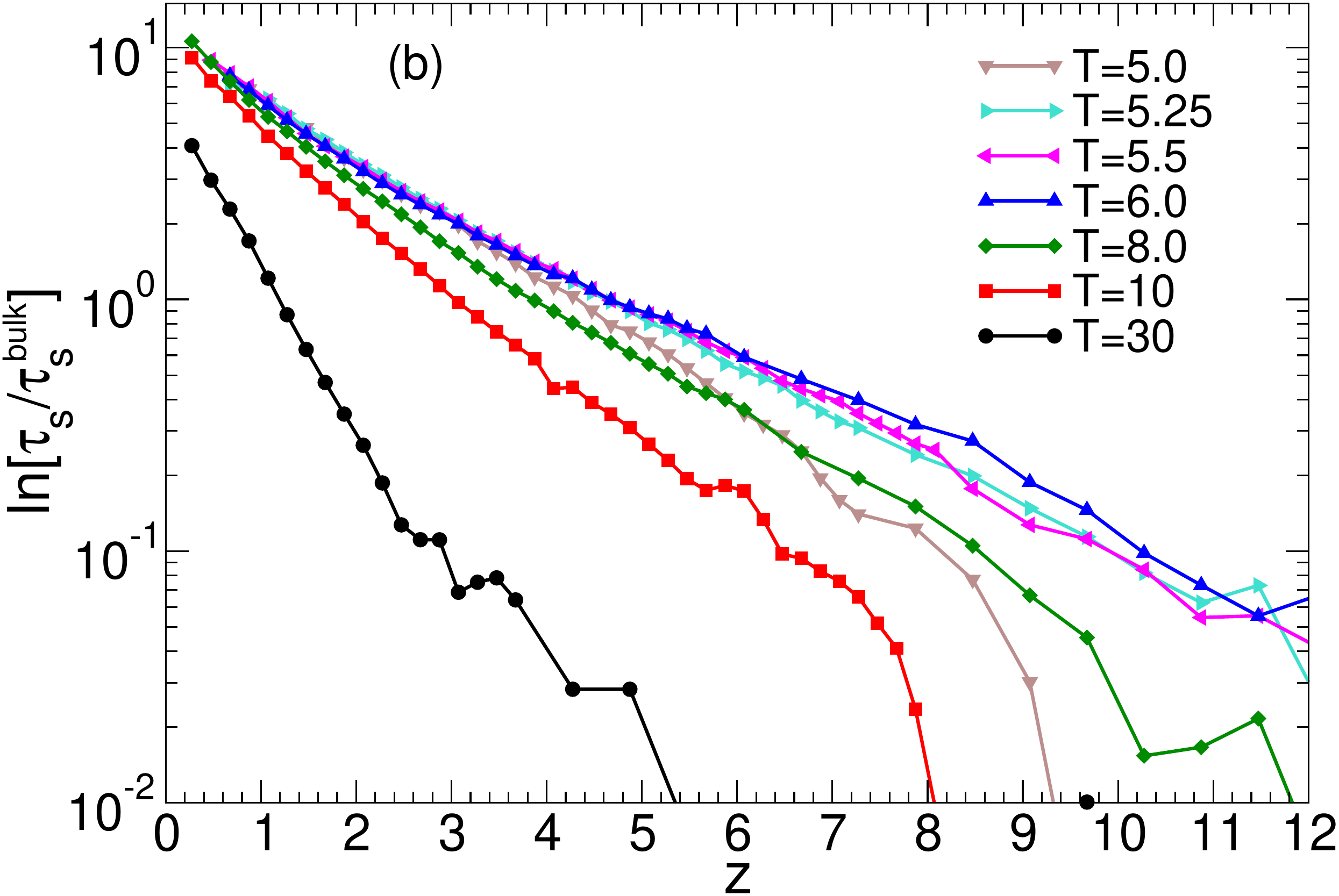}
\caption{a) $z-$dependence of the relaxation time as obtained for
the self-overlap. The different curves correspond to different
temperatures. b) Same data as in panel a) but plotted in a form to
show that at intermediate and large $z$ the data is compatible with the
functional form given by Eq.~(\ref{eq5}).}
\label{fig5}
\end{figure}

Before discussing the temperature dependence of these static length
scales we briefly turn our attention to the dynamics. From the fit
to the data with the functional form given by Eq.~(\ref{eq3}) we
can extract a relaxation time $\tau_c(z,T)$ {as well as its single
particle counterpart, $\tau_s(z,t)$, corresponding to the self overlap}.
In Fig.~\ref{fig5}a we show the $z-$dependence of $\tau_s(z,t)$, i.e.
the relaxation times extracted from the self-overlaps. (The data from
the collective functions look qualitatively similar). We see that
for all temperatures the relaxation times become independent of $z$
if $z$ is large which shows that the size of the simulation box is
sufficiently large to reach the bulk value.  For small distances from
the wall the relaxation times quickly increase and the $z-$range over
which this increase can be noticed increases with decreasing temperature.
This observation implies that there is a {\it dynamic} length scale that
growth if $T$ is lowered. Previous studies for a Lennard-Jones system
have shown~\cite{scheidler} that the $z-$dependence of $\tau_s(z,T)$
can be approximated well by

\begin{equation}
\log(\tau_s)= \log(\tau_s^{\rm bulk}) + B_s(T) \exp(-z/\xi_s^{\rm dyn}).
\label{eq5}
\end{equation}
 
\noindent
That this functional form does give a good description of the data also
for the present system is shown in Fig.~\ref{fig5}b. We see that for
intermediate and large values of $z$ the data points are indeed compatible
with a straight line, as expected from Eq.~({\ref{eq5}). Therefore we
can use this result to define a {dynamic} length scale $\xi_s^{\rm dyn}$.

A close inspection of the data shows that the slope of the curves at
intermediate and large $z$ does {\it not} evolve monotonically with
temperature. Instead the data for $T=6.0$ has the smallest slope and
the ones for $T=5.25$ and $T=5.0$ have a larger slope (in absolute
values).  This observation implies that the dynamic length scale is not a
monotonically increasing function of $T$ but instead shows a maximum. In
Ref.~\cite{kob_12} we have shown that this surprising result is not
just an artifact of our data analysis because {it can be seen} directly
in the time correlation functions and therefore must be considered
as a real effect. Finally we also note that for small values of $z$
the curves $\tau_s(z,T)/\tau_s({\rm bulk},T)$ become independent of
$T$. This implies that very close to the wall the temperature dependence
of the relaxation dynamics tracks the one of the bulk system. These
two results give indirect evidence that the relaxation dynamics occurs
on {\it two} different length scales: A first one that is relevant at
short scales and a second one, $\xi^{\rm dyn}$, that
operates at intermediate and long distances.  We will come back to this
interpretation below.

\begin{figure}
\hspace*{30mm}
\includegraphics[scale=0.32,angle=0,clip]{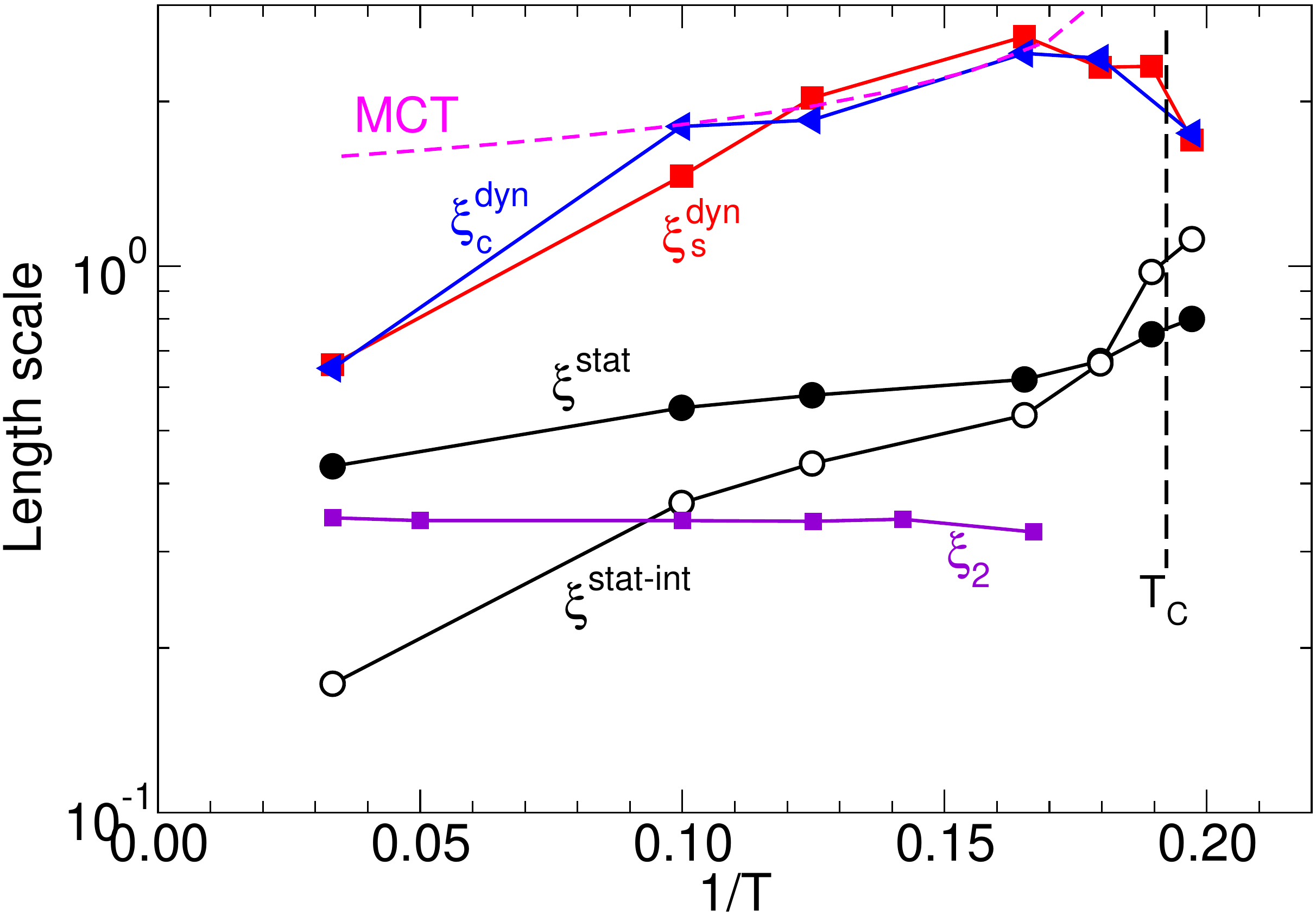}
\caption{Temperature dependence of different static and dynamic length scales.
The vertical dashed line indicates $T_c=5.2$.}
\label{fig6}
\end{figure}

In Fig.~\ref{fig6} we summarize the $T-$dependence of the various static
and dynamic length scales. The length scale labeled $\xi_2(T)$ has been
obtained by fitting the decay of the envelope of the radial distribution
function $g(r)$ for the larger particles to an exponential. As expected,
this length scale shows basically no $T-$dependence, in agreement
with the data shown in Fig.~\ref{fig1}. As discussed in the context
of Fig.~\ref{fig4}, our simulations allow to extract two static length
scales that are related to multi-point correlations: $\xi^{\rm stat}$
and $\xi^{\rm stat-int}$. The influence of the particles constituting
the wall is likely to be associated with $\xi^{\rm stat-int}$, but one
can expect that at very low temperatures (at present inaccessible to
computer simulations) the two scales coincide because the prefactor
$B(T)$ should saturate. From Fig.~\ref{fig6} we see that both scales
increase with decreasing temperature and that $\xi^{\rm stat-int}$
does show a stronger $T-$dependence than $\xi^{\rm stat}$. This latter
behavior is related to the fact that the prefactor $B$ in Eq.~(\ref{eq4})
also shows a significant $T-$dependence. From the figure we see that the
so defined static length scales are at intermediate and low temperatures
significantly larger than the one obtained from the two point correlation
function, thus showing that multi-point correlations are indeed of high
interest for glass-forming systems.

Also included in the figure are the dynamic length scales $\xi^{\rm
dyn}$ as determined from the collective and self overlaps. We
see that within the numerical noise of the data these two scales
coincides and that both of them are significantly larger than the
static ones. (Qualitatively similar results have been obtained for
different geometries of pinned particles, which indicates that this
result is quite general~\cite{berthier_12}.) The growth of the dynamic
length scales for $T>T_c$ (where the critical temperature $T_c$ from
the mode-coupling theory is indicated by a vertical dashed line) is
compatible with the prediction from mode-coupling theory~\cite{imct},
i.e. the functional form $\xi^{\rm dyn} \sim (T-T_c)^{-0.25}$ with
$T_c=5.2$ (dashed line). (See Ref.~\cite{kob_12} on how $T_c$ has been
determined.) But in view of the uncertainty in the value of $T_c$ and
the smallness of the exponent, this is not a very strong statement. More
important is the observation that these dynamic length scales show
a local maximum and that the temperature at which it occurs is close
to the critical temperature of mode-coupling theory.  This result is
thus {a} strong evidence that around $T_c$ the nature of the process
that is responsible for the relaxation dynamics changes. Note that,
in contrast to all previous investigations on the relaxation dynamics
of glass-forming liquids, the presence of the maximum is determined
without using any of the mode-coupling theory predictions. It represents
therefore a genuine physical phenomenon and not, for instance, deviation
from a fitting formula. We interpret this maximum as a {firm} evidence
that $T_c$ is indeed a {physically meaningful} crossover temperature
for the glass-former and not merely a fitting parameter.

\begin{figure}
\hspace*{37mm}
\includegraphics[scale=0.32,clip]{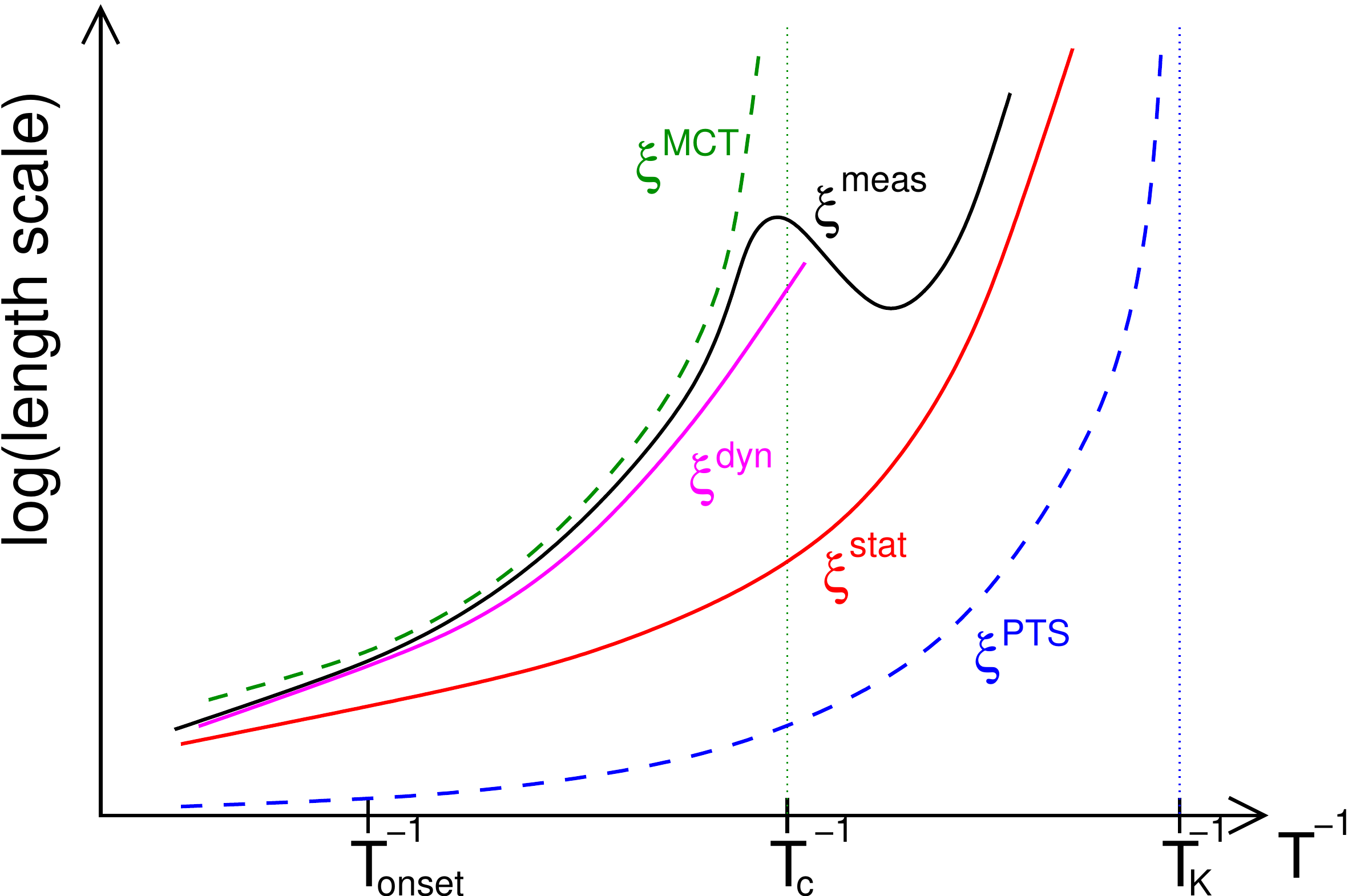}
\caption{Schematic plot of the temperature dependence of different static and 
dynamic length scales.}
\label{fig7}
\end{figure}

Although the present simulations do not really permit to draw
conclusions on the nature of the relaxation process at play in the various
temperature regimes, we can speculate about this issue.  The most natural
interpretation of these data is that dynamic heterogeneity, and therefore
structural relaxation, change nature when $T$ is decreased across the
mode-coupling crossover.  Based on the mean field picture, we suggest
the following scenario.  At high temperatures (i.e. in the normal liquid
state well above $T_c$) the particles relax in an independent manner and
the dynamic and static length scales are small. If $T$ is lowered to a
temperature at which the system starts to become sluggish (i.e. below a
temperature that is often called `onset temperature', $T_{\rm onset}$)
particle motion becomes more collective and spatially correlated, as
predicted for instance by mode-coupling calculations.  When temperature
is lowered towards $T_c$, however, two mechanisms start to compete. At
low temperature, dynamics becomes slower and spatially correlated over
a dynamic correlation length that keeps increasing, $\xi^{\rm dyn}$.
At the same time, close to $T_c$ static correlations also start to
become relevant and might control the dynamic profiles we observe very
close to the wall. Finally, when decreasing temperature below $T_c$,
the mode-coupling relaxation processes start to disappear ($T_c$ is
just a crossover), but static correlations remain and are now fully
responsible for the dynamic profiles.  In Fig.~\ref{fig7}, we represent
schematically the consequence of this scenario for length scales.  We show
both diverging length scales predicted within (mean field) RFOT, the
dynamic length scale of mode-coupling theory $\xi^{\rm MCT}$ at $T_c$,
and the static point-to-set length scale $\xi^{\rm PTS}$ at $T_K$.
In finite dimensions, $\xi^{\rm dyn}$ is controlled by $\xi^{\rm MCT}$
over the regime $T_{\rm onset} > T > T_c$, but this process progressively
disapppears near $T_c$.  Static correlations $\xi^{\rm stat}$, are
controlled by the point-to-set correlation functions.  As a result, the
measured dynamic lengthscale, $\xi^{\rm meas}$, is first controlled by
$\xi^{\rm dyn}$ and then by $\xi^{\rm stat}$ at lower temperature, and
thus $\xi^{\rm meas}$ exhibits a non-monotonic temperature dependence.
This scenario naturally explains the emergence of a non-monotonic
temperature evolution of the measured dynamic profile.  We note that
Stevenson {\it et al.} in Ref.~\cite{stevenson_06} have suggested a
similar scenario and have additionally proposed that the geometry of
spatial correlations changes from a fractal, string-like structure above
$T_c$, to more compact domains at lower temperatures. Our data are not
incompatible with these ideas, but they do not provide direct geometric
information about dynamic correlations.

\begin{figure}
\hspace*{30mm}
\includegraphics[scale=0.52,angle=0,clip]{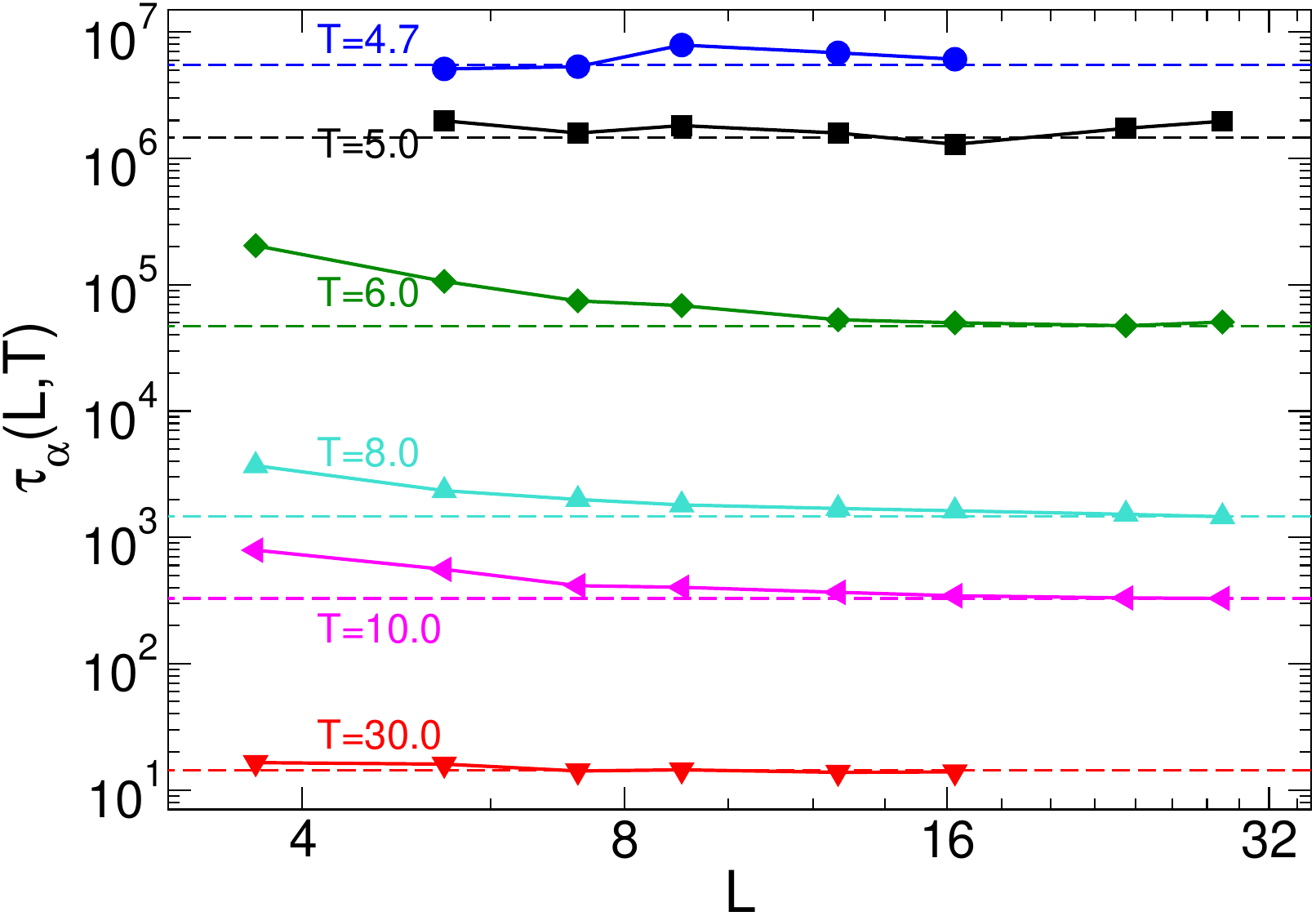}
\caption{Relaxation time as a function of the box size for different 
temperatures.}
\label{fig8}
\end{figure}

The above results have been  obtained for a very specific point-to-set
geometry using flat amorphous walls and one may wonder about their
relevance for bulk dynamics.  We now briefly discuss how the non-monotonic
$T-$dependence of $\xi^{\rm meas}$ can be used to understand some finite
size effects of the relaxation dynamics.  In Fig.~\ref{fig8} we show the
system size dependence of the relaxation time for different temperatures,
$\tau(L,T)$ for the same harmonic sphere system as the one studied near
the wall.  The relaxation time is defined as the time it takes for the
intermediate scattering function to decay to $1/e$ of its $t=0$ value.

The data show that at high temperatures there is no dependence on system
size $L$, which is reasonable since all the relevant length scales are
small. With decreasing $T$, i.e. $T=10.0$, 8.0, and 6.0, one finds that
$\tau$ increases if $L$ decreases, indicating that there are relaxation
processes that extend over scales that become larger and compete with the
finite system size. The temperature evolution of these data are compatible
with the existence of a growing dynamic length scale in this regime.
Interestingly, this effect is most pronounced for $T=6.0$, i.e. at the
temperature at which we find in Fig.~\ref{fig6} the maximum in $\xi^{\rm
dyn}$.  Remarkably, by decreasing further the temperature in the regime
where $\xi^{\rm dyn}$ was found to decrease, we find that the system size
dependence of the relaxation time becomes less pronounced. Finally, at
even lower temperature, $T=4.7$, the $L$ dependence of the relaxation
time qualitatively changes nature, in agreement with the idea that
dynamic process above and below $T_c$ are qualitatively different.
Thus, the non-monotonic temperature evolution of $\xi^{\rm meas}$ near
the amorphous wall is mirrored by a similarly non-monotonic temperature
evolution of the size dependence of the relaxation time in the bulk,
which we see as an independent confirmation of the speculative scenario
described by Fig.~\ref{fig7}.

\section{Summary}

The above results show that the point-to-set protocol of freezing a finite
set of particles in their equilibrium allows one to gain information
about relevant static and dynamic correlation functions in glass-forming
liquids. This approach opens the door to probe high order static and
dynamic length scales in a novel manner. We have found that there is
indeed a dynamic length scale that increases in qualitative agreement
with the mean field scenario, but that this scale decreases again if one
is significantly below the mode-coupling temperature. This result, which
has allowed to see a signature of $T_c$ without using any of the standard
fitting procedures, gives thus support to the theoretical calculations,
and suggests how the mean field predictions should be modified in finite
dimensions.  We hope that experiments will take up this approach and
therefore confirm the scenario we have developed here.
\\[5mm]

Acknowledgments: We thank G. Biroli and A. Cavagna for fruitful
exchanges about this work, and the R\'egion Languedoc-Roussillon (L.B.),
ANR DYNHET (L.B. and W.K.), and MICINN (Project: MAT2009-13155-C04-02)
and Junta de Andaluc\'{i}a (Project: P07-FQM-02496) (S.R.V.) for financial
support. W.K. is member of the Institut universitaire de France.
\\[5mm]





\bibliographystyle{elsarticle-num}
\bibliography{<your-bib-database>}

\begin{thebibliography}{00}

\bibitem{kob_97}
W. Kob, C. Donati, S.~J. Plimpton, S.~C. Glotzer, and P.~H. Poole,
Phys. Rev. Lett. {\bf 79}, 2827 (1997).

\bibitem{ediger_00}
M. D. Ediger,
{\it Annu. Rev. Phys. Chem.} {\bf 51} 99 (2000).

\bibitem{berthier_04}
L. Berthier,
Phys. Rev. E {\bf 69}, 020201 (2004).

\bibitem{book}
{\it Dynamical heterogeneities in glasses, colloids, and 
granular media}, Eds.: L. Berthier, G. Biroli, J.-P. Bouchaud,
L. Cipelletti, and W. van Saarloos, (Oxford University Press, 
Oxford, 2011).

\bibitem{berthier_05}
L. Berthier, G. Biroli, J.-P. Bouchaud, L. Cipelletti, D. El Masri,
D. L'H\^ote, F. Ladieu, and M. Pierno,
Science {\bf 310}, 1797 (2005).

\bibitem{dalle_07}
C. Dalle-Ferrier, C. Thibierge, C. Alba-Simionesco, L. Berthier,
G. Biroli, J.-P. Bouchaud, F. Ladieu, D. L'H\^ote, and G. Tarjus,
Phys. Rev. E {\bf 76}, 041510 (2007).

\bibitem{binder_kob_11}
K. Binder and W. Kob,
{\it Glassy materials and disordered solids, 2nd edition}
(World Scientific, Singapore, 2011).

\bibitem{BBepl}
G. Biroli and J. P. Bouchaud,
Europhys. Lett. {\bf 67}, 21 (2004).

\bibitem{silvio07}
S. Franz and A. Montanari, 
J. Phys. A: Math. Theor. {\bf 40}, F251 (2007). 

\bibitem{gotze_08}
W. G\"otze, {\it Complex dynamics of glass-forming liquids:
A mode-coupling theory} (Oxford University Press, Oxford, 2008).

\bibitem{imct}
G. Biroli, J.-P. Bouchaud, K. Miyazaki, 
and D. R. Reichman, Phys. Rev. Lett. {\bf 97}, 195701 (2006).

\bibitem{kirpatrick_89}
T. R. Kirkpatrick, D. Thirumalai,
and P. G. Wolynes,
Phys. Rev. A {\bf 40}, 1045 (1989).

\bibitem{xia_00}
X. Y. Xia and P. G. Wolynes,
Proc. Natl. Acad. Sci. USA {\bf 97}, 2990 (2000).

\bibitem{coslovich_07}
D. Coslovich and G. Pastore,
J. Chem. Phys. {\bf 127}, 124504 (2007).

\bibitem{tanaka}
H. Tanaka, T. Kawasaki, H. Shintani, and K. Watanabe,
Nature Mat. {\bf 9}, 324 (2010).

\bibitem{gilles}
G. Tarjus, S. A. Kivelson, Z. Nussinov, and P. Viot, J. Phys.
Condens. Matter {\bf 17}, R1143 (2005).

\bibitem{bb_04}
J.-P. Bouchaud and G. Biroli, 
J. Chem. Phys. {\bf 121}, 7347 (2004).

\bibitem{jack_05}
R. L. Jack and J. P. Garrahan,
J. Chem. Phys. {\bf 123}, 164508 (2005).

\bibitem{cavagna_07}
A. Cavagna, T. S. Grigera, and P.
Verrocchio, Phys. Rev. Lett. {\bf 98}, 187801 (2007).

\bibitem{biroli_08}
G. Biroli, J.-P. Bouchaud, A. Cavagna, T. S. Grigera, 
and P. Verrochio, Nature Phys. {\bf 4}, 771 (2008).

\bibitem{mosayebi_10}
M. Mosayebi, E. Del Gado, P. Ilg, and H. C. \"{O}ttinger, 
Phys. Rev. Lett. {\bf 104}, 205704 (2010)

\bibitem{kob_12}
W. Kob, S. Roldan-Vargas, and L. Berthier,
Nature Phys. 8, 164 (2012) 

\bibitem{berthier_12}
L. Berthier and W. Kob,
Phys. Rev. E {\bf 85}, 011102 (2012)

\bibitem{jack_12}
R. L. Jack and L. Berthier,
Phys. Rev. E {\bf 85}, 021120 (2012)

\bibitem{charbonneau_12}
B. Charbonneau, P. Charbonneau, and G. Tarjus,
Phys. Rev. Lett. {\bf 108}, 035701 (2012).

\bibitem{berthier_09}
L. Berthier and T. A. Witten,
EPL  {\bf 86}, 10001 (2009); Phys. Rev. E {\bf 80}, 021502 (2009).

\bibitem{scheidler}
P. Scheidler, W. Kob, and K. Binder,
Europhys. Lett. {\bf 59}, 701 (2002); J. Phys. Chem. B {\bf 108}, 6673 (2004).

\bibitem{berthier_tobe}
L. Berthier, G. Biroli, D. Coslovich, W. Kob and C. Toninelli,
arXiv:1203.3392.

\bibitem{stevenson_06}
J. D. Stevenson, J. Schmalian, and P.G. Wolynes
Nature Phys. {\bf 2}, 268 (2006).


\end{thebibliography}



\end{document}